# Nonvolatile Electrochemical Memory at 600 ºC Enabled by Composition Phase Separation

Jingxian Li[*,1], Andrew J. Jalbert[1], Leah S. Simakas[1], Noah J. Geisler[1], Virgil J. Watkins[1], Laszlo A. Cline[1], Elliot J. Fuller[2], A. Alec Talin[2], Yiyang Li[*,1]

[1]Materials Science and Engineering, University of Michigan, Ann Arbor, MI, USA

[2]Sandia National Laboratories, Livermore, CA, USA

*Corresponding authors: jxli@umich.edu, yiyangli@umich.edu

**Contact information:**
Jingxian Li, jxli@umich.edu
Andrew Jalbert: jalberta@umich.edu
Leah Simakas: lsimakas@umich.edu
Noah Geisler: geislern@umich.edu
Virgil J. Watkins: virgiljw@umich.edu
Laszlo Cline: lzcline@umich.edu
Elliot J. Fuller: ejfull@sandia.gov
A. Alec Talin: aatalin@sandia.gov
Yiyang Li: yiyangli@umich.edu

**Bigger Picture**

Moore's Law has led to monumental advances in computing over the past fifty years. However, one shortcoming of silicon-based logic and memory devices is their limited temperature range, typically <150°C. In this work, we present a solid-state memory device that can operate and store information at temperatures as high as 600°C. Rather than rely on the motion of electrons, this device stores information through the electrochemical migration of oxygen ions in transition metal oxides, a process that bear resemblance to solid oxide fuel cells and batteries. This memory device can expand the use of microelectronics in extreme environments like deep energy wells, turbine engines, and space exploration.

## Abstract

CMOS-based microelectronics are limited to ~150°C and therefore not suitable for the extreme high temperatures in aerospace, energy, and space applications. While wide bandgap semiconductors can provide high-temperature logic, nonvolatile memory devices at high temperatures have been challenging. In this work, we develop a nonvolatile electrochemical memory cell that stores and retains analog and digital information at temperatures as high as 600 °C. Through correlative electron microscopy, we show that this high-temperature information retention is a result of composition phase separation between the oxidized and reduced forms of amorphous tantalum oxide. This result demonstrates a memory concept that is resilient at extreme temperatures and reveals phase separation as the principal mechanism that enables nonvolatile information storage in these electrochemical memory cells.

## Introduction

Conventional Si-based digital electronics are typically limited to 150°C due to the rise in the intrinsic carrier concentration with temperature[1]. This limitation precludes high-temperature needs[2] including in aerospace and automotive engines (150-600°C), interplanetary exploration such as Venus (550°C), and the extraction of petroleum and geothermal energy from deep wells (300-600°C)[1]. Logic and memory form the foundation for modern computing. While high-temperature logic can be developed using wide bandgap semiconductors like Silicon Carbide (SiC)[3,4] and Gallium Nitride (GaN)[5], as well as emerging technologies like carbon nanotubes[6], the development of memory units for such extreme temperatures has been more challenging. Presently, the only re-writeable memories[7] that have retained nonvolatile information above 500°C are nanogap resistive memory[8] and nitride-based ferroelectric memory[9,10]; these emerging technologies exhibit limitations including large cycle-to-cycle variations, destructive read, and/or high switching voltages.

Electrochemical Random-Access Memory (ECRAM) has recently emerged as a promising solution for analog computing[11–15]. ECRAM is a three-terminal memory device that stores and switches information through ion migration and the resulting change in valence and electronic conductivity. ECRAM research primarily focused on three types of ions: $Li^+$ [16–20], $H^+$ [21–25], and $O^{2-}$ [26–36]; other ions like the $Mg^{2+}$ [37] have also been studied. Oxygen-based ECRAM operates by electrochemically transporting oxygen ions (or vacancies) between two mixed ionic and electronic conducting (MIEC) metal oxides with a solid electrolyte. This mechanism allows ECRAM to switch resistance states by modulating the oxygen concentration in the MIEC metal oxides. Early demonstrations of oxygen ion ECRAM based on titanium[27] or tungsten[26] oxides achieved a high density of analog states but suffered from poor state retention times (<1 day at room temperature). Recently, it was shown that tungsten oxide-based ECRAM can yield ~24 hours of retention at 200°C under short circuit[38], which projects to over 10 years at 85°C. However, this device was constructed on single-crystal yttrium stabilized zirconia (YSZ) crystal substrate and therefore not CMOS compatible. Moreover, there is no clear pathway for operations above 200 °C.

In this work, we developed a high-temperature stable ECRAM device utilizing amorphous $TaO_x$ deposited on a $Si/SiO_2$ substrate that is capable of recording and retaining binary and multi-level analog conductance states at temperatures up to 600 °C. Importantly, we obtain over 24 hours of retention at short circuit at 600°C, a notable advancement in the resilience of electrochemical memory devices. We propose that these findings are a result of amorphous phase separation between the Ta and $TaO_{1.9}$[39,40] which have been shown to facilitate information retention in $TaO_x$-based resistive memory devices[40]. Our correlative transmission electron microscopy shows clear evidence of this phase separation. Our results demonstrate a promising path towards the realization of high-temperature analog and binary resistive memory.

## *Results*

**TaOx-based ECRAM as High-Temperature Nonvolatile Binary Memory**

The structure of our $TaO_x$-ECRAM cells consists of a 20 nm reservoir layer of tantalum, a 140 nm electrolyte layer of yttria stabilized zirconia (YSZ) electrolyte, and a 20 nm channel layer of amorphous oxygen-deficient tantalum suboxide ($TaO_x$) deposited on $SiO_2$ (500 nm)/Si substrates using sputtering (Fig. 1a, S1). Three Pt current collectors are sputtered using shadow masks to serve as the gate, source, and drain electrodes. Finally, a passivation layer of $SiN_X$ is deposited to protect the tantalum oxide from moisture and oxygen in the environment. We switch this device by applying either a positive or a negative gate voltage ($V_G$), typically +2V to increase (SET) or -2V to decrease (RESET) the channel conductance. During retention measurements, we ground the gate by applying 0 V, short-circuiting it to the channel. No switch or selector was used in these experiments. All fabrication was conducted at room temperature with a 1 hour anneal 400°C to improve the crystallinity of the YSZ but still maintain thermal compatibility with back-end-of-line processes. More details are in the experimental section.

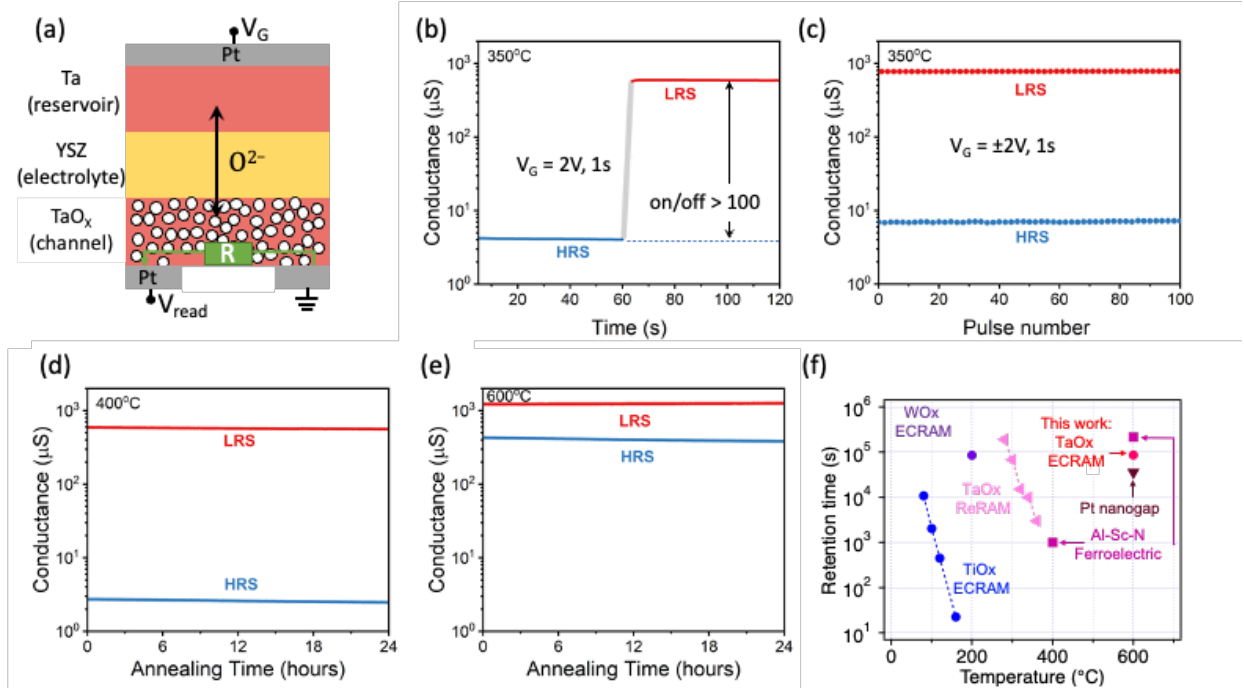

**Figure 1. Binary switching and high-temperature retention of TaOx ECRAM devices.** (a) Schematic depicting the structure of the ECRAM device, featuring a stacked configuration of $TaO_x$/YSZ/Ta. This design utilizes an electric field to facilitate the movement of oxygen ions through the YSZ electrolyte, effectively modulating the oxygen concentration in the $TaO_x$ channel oxide to enable the storage of information through the creation of varying resistance states in the device. (b) Binary switching behavior from HRS to LRS with one pulse (2V, 1s) at 350°C shows an on/off ratio of more than 100. (c) 50 cycle switching between LRS and HRS with minimal cycle-to-cycle variation. (d-e) The retention measurements of two resistance states at 400°C (d) and 600°C (e) for 24 hours without failure. At 400°C, the LRS conductance decreased by 5%, and the HRS conductance decreased by 10% after 24 hours; at 600°C, the LRS conductance increased by 5% while the HRS conductance decreased by 12% after 24 hours. (f) Retention times of various memory devices including our ECRAM with $TaO_x$, Pt nanogap devices[8], $Al_{0.7}Sc_{0.3}N$ ferroelectrics[9,10], ReRAMs based on $TaO_x$[41], and ECRAMs utilizing $TiO_x$[27] and $WO_x$[38].

We initially assess the binary switching behavior. Fig. 1b illustrates the SET process, showcasing a >100 on/off ratio differentiating between the high resistance state (HRS) and low resistance state (LRS) with ± 2V for 1 second at 350°C. This SET/RESET switching is reproducible (Fig. 1c) with minimal cycle-to-cycle variation. Fig. S2 shows that resistance switching could be effectively obtained across a broad temperature range of 200°C to 600°C, rendering this technology adaptable for diverse high-temperature applications. Moreover, the endurance exceeds 10000 pulses at 400°C (Fig. S3). To assess the temperature-dependent switching, we further conducted a comprehensive examination of temperature's influence on switching behavior by utilizing a consistent 2V, 1s pulse across a temperature span from 300°C to 550°C (Fig. S4). Our findings reveal that the highest on/off ratios (>100) are obtained for temperatures ranging between 350°C and 400°C (Fig. S5). However, at temperatures below 300°C, the sluggish oxygen diffusivity in our metal oxides impedes sufficient ion migration needed for higher on/off ratios. At temperatures above >450°C, the HRS conductance increases due to an increase in the electronic conductivity of $TaO_2$ with temperature, and thus reduces the ON/OFF ratio.

We tested the retention of the device under short circuit at temperatures between 400°C to 600°C. These results show less than 12% change in the conductance after 24 hours over these temperatures (Fig. 1d-e, Fig. S6). Fig. 1f shows a comparative analysis of the retention times between our $TaO_x$-ECRAM and other high-temperature memory technologies (Fig. 1f). At 600°C, our ECRAM cell has similar temperature performance to Pt-nanogap ReRAM[8] and $Al_{0.7}Sc_{0.3}N$ ferroelectric capacitors[9]; a comparative analysis of these devices will be given in the discussion section. ECRAMs utilizing $TiO_x$[27] and $WO_x$[38] show substantially reduced retention compared to the $TaO_x$ ECRAM device shown here. Likewise, filament-based ReRAM devices using similar materials[42] yield substantially lower retention compared to the $TaO_x$ ECRAM cell (Fig. 1f).

**Direct Visualization of Phase Separation in $TaO_X$ ECRAM**

To shed light on the mechanism of the above resistive switching and retention behaviors in $TaO_x$-based ECRAM, we performed Scanning Transmission Electron Microscope (STEM) on cross-sectioned devices. Fig. 2a depicts high-angle annular dark field (HAADF) and energy dispersive spectroscopy (EDS) mapping at the Ta K edge and O K edge from a pristine Ta/YSZ/$TaO_2$ device. Although the EDS atomic ratios are not fully quantitative, they are consistent with a metallic Ta ion reservoir and a $TaO_2$ channel based on our deposition recipes. Next, we consider a device after switching to the LRS in Fig. 2b. On the channel side, we observe a ~5 nm reduced Ta-rich layer at the electrolyte/channel interface and a corresponding ~15nm oxidized $TaO_X$ at the reservoir/electrolyte interface. These results are consistent with the migration of oxygen ions from the channel layer through the electrolyte in response to the applied electric field, ultimately reaching the Ta reservoir. The reduced Ta-rich layer at the channel is consistent with the increased conductance of the low-resistance state. The increased thickness of the reservoir may be attributed to the volumetric expansion due to the oxidation of the Ta to $TaO_x$, while the reduced channel undergoes slight contraction due to the loss of oxygen ions. Our STEM results provide direct microscopic evidence to show the electrochemical migration of oxygen in ECRAM devices.

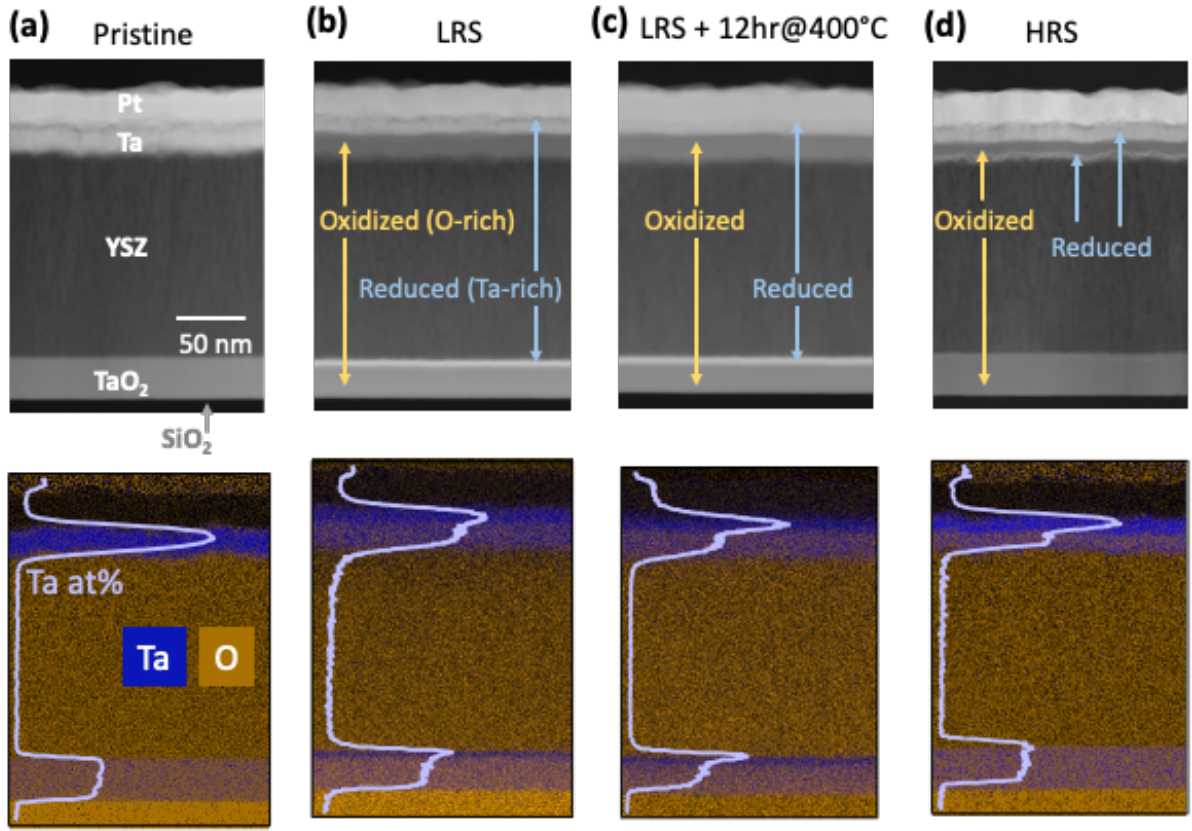

**Figure 2. Scanning Transmission Electron Microscopy (STEM) of cross-sectioned ECRAM devices.** The top row is the high-angle annular dark field (HAADF) mapping, while the bottom row is energy dispersive spectroscopy (EDS) mapping. The Ta at% is plotted as a function of the depth. (a) Pristine devices fabricated using a Ta ion reservoir and a $TaO_2$ channel; this sample was not annealed at 400°C after fabrication. (b) In the low-resistance state (LRS), the ion reservoir is partially oxidized while a reduced Ta-rich, O-poor layer is formed at the channel. (c) The LRS device maintains its metal and oxygen distribution even after short-circuiting the device for 12 hours at 400°C. This coexistence of oxidized and reduced $TaO_x$ compounds over time in both the reservoir and the channel is consistent with the composition phase separation of the material and the nonvolatile information retention of the device. (d) The HRS state shows that the channel is again oxidized, while a reduced Ta-rich layer forms at the reservoir layer.

We next investigate retention by imaging a different LRS device after 12 hours of short circuit at 400°C (Fig. 2c). The tantalum and oxygen distribution remains essentially unchanged compared to that observed in the LRS device without this high-temperature annealing (Fig. 2b). This observation is consistent with the stability of the electrical resistance over this time (Fig. 1d). We attribute this stability to the phase separation of the tantalum and oxygen within the device, which appears to play a crucial role in retaining the chemical state and, by extension, the electrical properties under thermal stress (Fig. 1d-e). Specifically, within the miscibility gap, the oxygen chemical potential of all compositions of Ta and O is equal (Fig. S7). For this reason, the coexistence of oxidized and reduced layers in both the channel and the reservoir is a stable configuration; it will not revert to a single homogeneous film over time. For the same reason, the oxygen chemical potential in the reservoir and channel are equal regardless of the thickness of the reduced Ta-rich layer in the channel. As a result, oxygen has no preference to migrate from the gate to the channel or vice versa, resulting in the retention stability of the device in Fig. 1d. In other words, the long retention of our ECRAM device at high temperatures under short circuit is enabled by the absence of any chemical driving forces to move the oxygen ions between the reservoir and channel.

In Fig. 2d, we image the high-resistance state after resetting the device. The channel composition resembles the pristine device in Fig. 2a, but with a slight increase in the Ta concentration. This likely accounts for the higher conductance seen in HRS (~3 μS at 400°C) relative to the pristine configuration (< 0.1 μS). Conversely, the HRS shows a reduced Ta layer at the reservoir-electrolyte interface and an oxidized layer above it, distinguishing it from both the pristine and the LRS state. We hypothesize that there is a net transfer of oxygen from the electrolyte to the metallic reservoir during the initial heating process, thereby making the reservoir in the HRS more oxidized than in the pristine device.

To further illustrate the role of phase separation, we conduct cyclic voltammetry between the reservoir/gate and the channel. The cyclic voltammetry in Fig. 3a shows clear oxidation and reduction peaks corresponding to the SET and RESET of the device. When a positive voltage is applied to the gate, the positive gate current indicates that negatively charged oxygen ions are moved from the channel to the ion reservoir. This results in the formation of a reduced Ta-rich layer in the channel, increasing the channel conductance (Fig. 3b). However, upon further increase of the voltage, the gate current decreases substantially, which most likely suggests a depletion of oxygen from the channel. A similar behavior is observed upon reversing the voltage, whereby a negative current implies that oxygen moves from the gate to the channel (Fig. 3a), reducing the channel conductance (Fig. 3b). This result is largely repeatable from one cycle to the next.

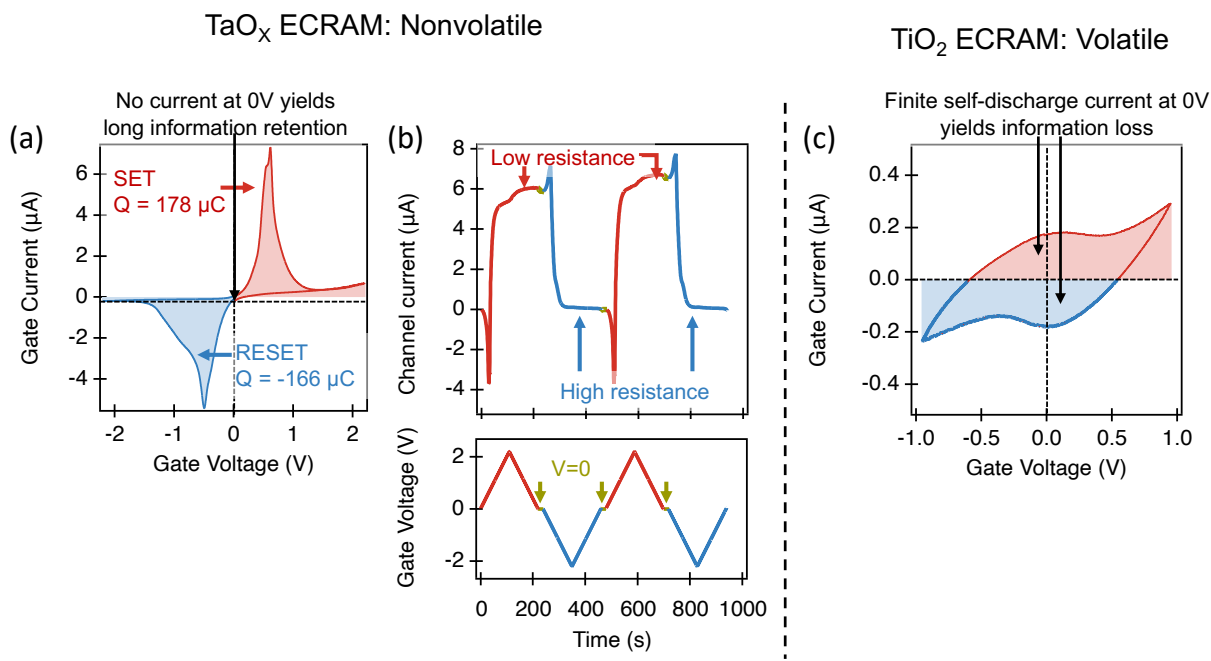

**Figure 3: Cyclic Voltammetry Analysis of ECRAM device.** (a) Cyclic voltammetry conducted between the gate and the channel shows clear redox peaks during SET and RESET. There is no current at 0V, consistent with the absence of an oxygen chemical potential difference between the reservoir and the channel. (b) The channel current as measured with a 10-mV potential taken concurrently with the gate cyclic voltammetry. During switching (SET or RESET), the channel current is affected by the electrochemical gate current from (a). During the 20-second OCV (yellow), we see the change in the resistance. (c) Cyclic voltammetry data for a volatile $TiO_2$-based ECRAM devices from previous work[27] shows a substantial negative current at 0V after SET, thereby resulting in information loss through an electrochemical self-discharge process.

The total integrated charge is about 170 micro-Coulombs in each direction, or $5.3 \times 10^{14}$ oxygen ions. With an estimated channel volume of $1.7 \times 10^{13}$ nm$^3$ (500 μm × 1750 μm × 20 nm), our results suggest that 30 oxygen ions are transferred for every (nm)$^3$ of device volume. Crystalline $Ta_2O_5$ has about 50 oxygen ions per (nm)$^3$ based on a density of 7.8 g cm$^{-3}$ and a molar mass of 442 g mol$^{-1}$. If the amorphous $TaO_2$ channel has a similar volume density of oxygen as bulk crystalline $Ta_2O_5$, the total charge transferred in the cyclic voltammetry suggests that most of the available oxygen has been transferred between the channel and ion reservoir. In contrast, our STEM images only show a partial reduction of the channel (Fig. 2b) because the switching was only conducted for 1 second.

The cyclic voltammetry sweeps also help us understand retention mechanisms. There is no electrochemical current at 0V, regardless of whether the channel was previously oxidized or reduced. This indicates that there is no migration of oxygen between the gate and the channel at 0V and is consistent with the nonvolatility observed at high temperatures (Fig. 1). This result is again consistent with phase separation whereby the chemical potential of the channel and reservoir are equal, thereby resulting in no electrochemical current when the voltage equals 0. In contrast, previous cyclic voltammetry on a $TiO_2$ ECRAM device shows a more capacitive profile whereby there exists a substantial current at 0V (Fig. 3c). This nonzero current reverses the oxygen migration forced at positive and negative voltages, thereby resulting in a loss of information over time.

**Analog Multilevel Switching**

In addition to binary switching, our $TaO_X$-ECRAM cells exhibit analog switching characteristics, which align with the traits observed in previous room-temperature ECRAM cells[11]. This device attains and sustains 100 analog states through sub-millisecond pulses at 400°C, as illustrated in Figure 4a. These properties mirror those found in other ECRAM and highlight the device's pivotal ability for analog matrix multiplications for in-memory computing. Moreover, our devices show reproducible switching consistent with previous oxygen-based ECRAM cells[26–28,43]. Fig. S8,9 shows the ECRAM device manifests analog behavior across different resistance values spanning from the micro-Siemens to milli-Siemens range This reliability and consistency further substantiate the device's potential for dependable analog data manipulation in various computational tasks.

We next discuss how the programming pulse characteristics affect switching. In probing the analog switching behavior across pulse widths ranging from 500 μs to 4 ms at 8V (Fig. S10a), we observe linear dependence of analog switching with pulse width. A similar behavior has been observed with a negative pulse (Fig. S10b). This indicates that the device's conductance can be finely tuned by altering the duration of the applied pulses. Similarly, Fig. S11 reveals that the conductance change is also linearly proportional to the pulse amplitude, consistent with previous reports in ECRAM[27].

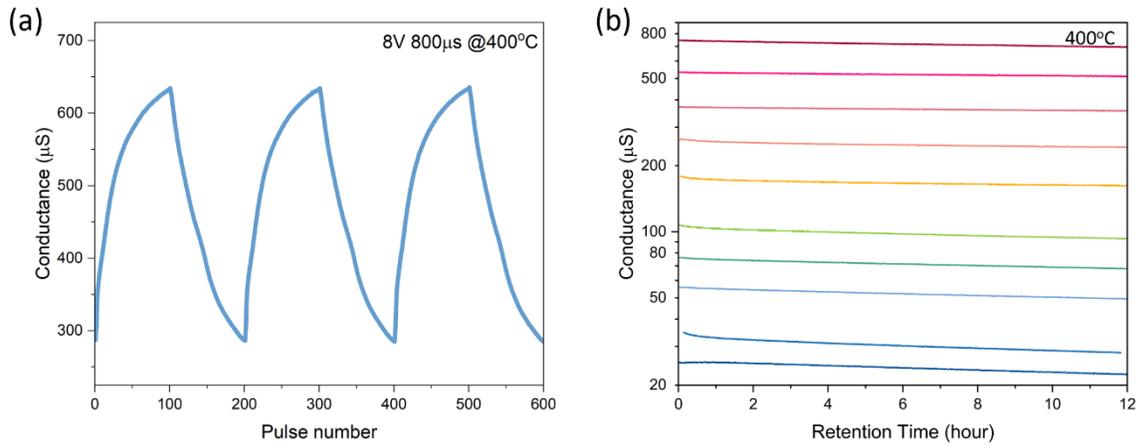

**Figure 4. Analog switching and retention characterizations of the ECRAM.** (a) Linear and symmetric conductance update in an ECRAM using a series of SET/RESET pulsing cycles of different pulse widths at 400°C (100 pulses, ±8 V, 800 μs). (b) Retention test of >10 analog states over a wide switching window.

Finally, we conduct data retention tests for the analog states, summarized in Fig. 4b. These tests show that the analog states ranging from 20 μS to over 800 μS maintained stable resistances after 12 hours at 400°C. The observed phase separation within our device contributes to establishment of multiple equilibrium resistance states.

## *Discussion*

Our study presents a non-volatile ECRAM device concept based on amorphous, phase separating $TaO_x$, demonstrating record ECRAM state retention in both binary and analog switching modes at temperatures as high as 600°C. The high-temperature performance is substantially better than commercial floating-gate memory (<150°C). It is also significantly better than all previously reported ECRAM types, which are limited to 200°C. Its temperature performance is comparable to that of recently demonstrated nitride-based ferroelectric memory[10] and Pt nanogap memory[8]. Compared to nitride ferroelectric memory, $TaO_x$ ECRAM uses lower voltages (2V vs >15V), can provide analog multi-level states, and has a higher ON/OFF ratio (100 vs 3 at 400°C, 3 vs 1.5 at 400°C). Furthermore, its resistance information state can be "read" without disturbing the device. Compared to the nanogap memory, ECRAM has lower voltages (2V vs 6V) and analog multi-level state capability (Fig. 4), which can result in a higher density of information states. We note that these three emerging memory technologies utilize very different switching and information retention mechanisms; each device has its advantages and disadvantages, and are all viable candidates for high-temperature nonvolatile memory.

One weakness of our ECRAM cell is the high write temperatures above 250°C. For low-temperature operations, we envision that it will be necessary to integrate a localized Joule heater as recently demonstrated for other oxygen-based ECRAM devices[44,45]. Additionally, while the ON/OFF ratio is diminished at 600°C, this is a result of an increased conductance of the channel layer and could be mitigated by making the channel more oxidized initially.

Our STEM and electrochemical results are consistent with an amorphous phase separation mechanism to explain the long-term retention in the $TaO_x$ ECRAM. This design rule of using composition phase separation can also be used to engineer other types of ECRAM, such as protonic ones which have fast speed but limited retention under short circuit[23,46]. While our previous work postulated that phase separation enables information retention in $WO_X$ ECRAM[38], this work shows much higher operations (600°C vs 200°C). We also directly characterized this phase separation using STEM (Fig. 2). Moreover, the persistent coexistence of the Ta-rich and oxygen-rich regions provides additional evidence of phase separation in amorphous tantalum oxide within the device. Composition phase separation has been shown to provide a path for information retention in both tantalum oxide resistive memory[40] as well as a lithium-based resistive memory based on lithium titanium oxide[47,48]. Due to the bulk dimensional switching region, our characterization of phase separation in ECRAM could also help the understanding of filamentary valence-change memory made using tantalum oxide and other materials, where the nanosized filaments are much more difficult to locate.

Finally, we speculate on the possible ultimate switching temperature of this device. Tantalum oxide is also known to crystallize around 650°C[49]; this could represent the upper temperature limit. However, if we can make ECRAM using crystalline tantalum oxide, then it could point to much higher operational temperatures. According to the Ta-O phase diagram[50], the miscibility gap extends to the melting temperature that exceeds 1500°C. In this regime, the functionality of the other components, such as the $SiN_X$ oxygen diffusion barrier, may limit the ultimate temperature limit.

## *Conclusion*

We present a nonvolatile electrochemical memory that switches and retains information at 600°C for over 24 hours. Our electrochemistry and transmission electron microscopy results suggest a phase separation mechanism is occurring in the amorphous tantalum oxide channel. This work shows new concepts that can be used to engineer nonvolatile memory for extreme environments.

## *Experimental Section*

**Fabrication of $TaO_X$ ECRAM Cells**

The ECRAM devices comprising thin-film YSZ were fabricated using an AJA Orion 8 DC/RF sputter system (baseline pressure <5 e-7 torr) using shadow masks on a 1 cm die of Si with 500 nm of thermal oxide (University Wafers). The first shadow mask (Fig. S1a) defines the bottom Pt electrode, which contains 5 nm of Ta adhesion layer and 20 nm of Pt. The Ta adhesion layer was DC sputtered at 100W on a 3-inch Ta target (99.95% purity; Plasmaterials Inc) under pure Ar (6N purity). The Pt layer was DC sputtered at 100W on a 2-inch Pt target (99.99% purity, AJA) under pure Ar. The sputter gas pressure was 3 mtorr.

The second shadow mask (Fig. S1b) defines the three successive oxide layers: the $TaO_x$ channel, the YSZ electrolyte, and the Ta reservoir layer. The channel, or switching layer, consists of 20 nm of $TaO_x$ and was DC sputtered at 100 W with the 3-in. Ta target and a sputter gas mixture of Ar (6N):$O_2$ (5N) at a 37:3 ratio regulated by mass flow controllers. The sputter gas pressure was maintained at 5 mtorr, with a substrate-to-target distance of approximately 15 cm. The 140 nm YSZ electrolyte was grown by RF sputtering on a 3-in. YSZ target (8 mol% $Y_2O_3$ in $ZrO_2$, 99.9% purity, Plasmaterials Inc) using a sputter power of 150 W and 5 mtorr of pure argon. The 20 nm Ta gate layer was grown by DC sputtering on a 3-in. Ta target at a sputter power of 100 W in 5 mtorr of pure argon.

The third shadow mask (Fig. S1c) was used to define the top gate contacts. It consists of 30-nm-thick gate Pt current collectors sputtered using the same process as the bottom contact, except without an adhesion layer. No additional annealing was performed during the room temperature sputtering process. Finally, for environmental protection against oxidation, a 60 nm silicon nitride layer was deposited on the ECRAM device's channel side using plasma-enhanced chemical vapor deposition (PECVD) at 200°C in the Lurie Nanofabrication Facility via a Plasmatherm 790. The $SiN_x$ above the current collectors were etched using reactive ion etching used a gaseous $CF_4/O_2$ mixture at a ratio of 95:5. The final devices are given in Fig. S1d,e. A 400°C anneal at 1 hour was conducted in inert Ar to improve the crystallinity and ionic conduction of the YSZ electrolyte.

**Device Measurements**

Following fabrication, the cell was tested in a six-probe Nextron MPS-Ceramic Heater CHH750 probe station, where the two channel electrodes made direct contact with Rh probes. Three of the probes were used to contact the top electrode and the two bottom electrodes. The environment was further regulated by flowing 5N ultra high-purity Ar at a controlled rate of 87 sccm using an Omega mass flow controller. A Swagelok check valve maintained the chamber pressure at ≈50 torr above ambient, further preventing air backflow. Monitoring with a Zirox ZR5 oxygen sensor indicated an Ar flow gas oxygen concentration of about 3 parts per million. Notably, the flowing Ar formed a more effective thermal contact between the heater and chip compared to testing the device in ultra-high vacuum. Device measurements were conducted utilizing a Bio-logic SP300 bipotentiostat or a National Instruments Data Acquisition Device (DAQ-6358) system.

**STEM Measurements**

STEM analyses were carried out using a Thermo Fisher Talos F200X G2, a 200 kV FEG scanning transmission electron microscope operating in STEM mode. Acquisition of STEM images and EDS data employed Velox software. STEM specimens were prepared utilizing a Thermo-Fisher Helios G4 Plasma FIB, with the final beam condition set at 12 keV 10 pA for liftout polishing.

**Acknowledgements**

The work at the University of Michigan was supported by the National Science Foundation under grant no. ECCS-2106225 and CCF-2235316, and by a Sandia University Partnership Network (SUPN) program. The authors acknowledge the Michigan Center for Materials Characterization for the use of the instruments and staff assistance. Part of the fabrication of the devices was conducted at the University of Michigan Lurie Nanofabrication Facility.

The work at Sandia National Laboratories was supported by the Laboratory-Directed Research and Development (LDRD) program. Sandia National Laboratories is a multi-mission laboratory managed and operated by National Technology and Engineering Solutions of Sandia, LLC, a wholly-owned subsidiary of Honeywell International Inc., for the U.S. Department of Energy's National Nuclear Security Administration under contract DE-NA-0003525.

**Author Contributions**

J.L., A.A.T., E.J.F., and Y.L. conceived the idea. J.L. designed the experiments. J.L., A. J. J., L.S., N. J. G., V. W., and L. C. fabricated ECRAM devices and performed device measurements. J.L. conducted transmission electron microscopy experiments. Y.L. supervised the project. All authors contributed to writing or revising the manuscript.

**Competing Financial Interests**

The authors have no financial interests to declare

**Declarations of Interest**

The authors have filed a patent based on this work to the US Patent and Trademark Office, application number 18/ 775896

**Data Availability**

The data that support the findings of this study are openly available at Materials Commons 2.0 at https://doi.org/10.13011/m3-pmdj-mz08